# High temperature onset of field-induced transitions in the spin-ice compound $Dy_2Ti_2O_7$

M. J. Matthews,[a] C. Castelnovo,[b] R. Moessner,[c] S. A. Grigera,[d,e] D. Prabhakaran,[f] and P. Schiffer[g]

[a] *Department of Physics and Materials Research Institute, Pennsylvania State University, University Park, PA 16802, USA*

[b] *South East Physics Network and Hubbard Theory Consortium, Department of Physics, Royal Holloway University of London, Egham TW20 0EX, UK*

[c] *Max-Planck-Institut für Physik komplexer Systeme, Dresden, 01187, Germany*

[d] *SUPA, School of Physics and Astronomy, University of St Andrews, St Andrews KY16 9SS, UK*

[e] *Instituto de Física de Líquidos y Sistemas Biológicos, UNLP-CONICET, La Plata 1900, Argentina*

[f] *Department of Physics, Clarendon Laboratory, University of Oxford, Parks Road, Oxford OX1 3PU, UK*

[g] *Department of Physics, University of Illinois at Urbana-Champaign, Urbana, IL 61801-3080*

We have studied the field-dependent ac magnetic susceptibility of single crystals of $Dy_2Ti_2O_7$ spin ice along the [111] direction in the temperature range 1.8 K - 7 K. Our data reflect the onset of local spin ice order in the appearance of different field regimes. In particular, we observe a prominent feature at approximately 1.0 T that is a precursor of the low-temperature metamagnetic transition out of field-induced kagome ice, below which the kinetic constraints imposed by the ice rules manifest themselves in a substantial frequency-dependence of the susceptibility. Despite the relatively high temperatures, our results are consistent with a monopole picture, and they demonstrate that such a picture can give physical insight to the spin ice systems even outside the low-temperature, low-density limit where monopole excitations are well-defined quasiparticles.

The geometrically frustrated spin ice materials have generated considerable attention for their exotic low temperature magnetic behavior. These materials, with the generic formula of $RE_2TM_2O_7$ (where RE is Ho or Dy, and TM is Ti or Sn), have a pyrochlore structure, with the magnetic rare earth ions located on a lattice of corner-sharing tetrahedra. The crystal field structure of these materials leads to a magnetic ground state doublet for the rare earth ions, with the first excited state separated from the ground state by a few hundred Kelvin. At low temperature, this results in effectively Ising spins with moments aligned along the local <111> crystalline direction, pointing along the line joining the centers of contiguous tetrahedra. Exchange and dipolar interactions, which have an energy scale of order 1-2 Kelvin, favor the "ice rules" stating that on each tetrahedron, two spins should point in and the other two out. Crucially, enforcing these rules across the entire system still allows an effective zero point entropy[1,2], and the ensemble of the exponentially many states obeying the ice rules underpin the novel physics found in spin ice. In particular, recent work has suggested that the elementary excitations out of this novel spin ice state have the characteristics of magnetic monopoles[3].

While the lowest temperature behavior of the spin ice materials has drawn the most attention, the dynamics of the spins at relatively high temperatures have also been explored by μSR[4,5], neutron spin echo[6-11], and ac magnetic susceptibility[2,12-16]. These studies have shown a range of interesting behaviors associated with the thermal and quantum relaxation of the spins and with the onset of spin ice order. Notably, while spin relaxation at higher temperatures is thermally activated, in the range of $T$ ~ 3K - 13K there is evidence for a temperature-independent (quantum) spin relaxation process[6,8,9,16].

Here we study the ac magnetic susceptibility of single crystals of the canonical spin ice compound $Dy_2Ti_2O_7$ in the presence of a magnetic field along the [111] crystal axis direction, which induces a quasi-two-dimensional "kagome ice" state at intermediate fields[17]. In particular, we explore the temperature regime near the onset of local spin ice correlations, in which there is a large population of thermal excitations. We find that, even though monopole excitations cannot be considered as well-defined quasiparticles at these relatively high temperatures, the monopole picture can nonetheless be helpful in clearly identifying the qualitative features of the regimes that emerge as the ice rules start asserting themselves at temperatures of order 2 K[2, 12, 18]. We also observe features associated with kinetic constraints in the kagome-ice state, presaging the dynamical arrest observed in spin ice at low temperatures[13, 16, 19]. In addition, we find a surprising temperature-independence in the susceptibility at an intermediate field in the kagome-ice regime, suggesting the possibility of more complex quantum effects than have been previously considered.

External fields applied along the [111] crystal axis direction of spin ice materials are of particular interest due to the geometry of the pyrochlore lattice and the strong spin anisotropy. The rare-earth sublattice can be viewed as stacked layers of triangular and kagome planes normal to the [111] direction, as indicated in Figure 1a. For each tetrahedron, three of the spins lie in a kagome plane, and the remaining spin lies in a triangular plane. The local Ising axes of spins in the triangular planes are aligned perpendicular to the planes (i.e., along the field direction)[20], while the spins in the kagome planes are at approximately 19.5º above or below the plane. With increasing field, the triangular plane spins are aligned the most easily with the field, because of the larger Zeeman interaction. An applied magnetic field of moderate strength (of order 0.3 T for T ~ 2 K) can effectively align the triangular spins without violating the local ice rules (Figure 1b), leaving the kagome plane spins in a disordered state known as "kagome ice"[21, 22]. For a sufficiently high field, all the spins on the kagome plane orient to have a positive projection onto the field direction, resulting in a 3-in/1-out or 3-out/1-in arrangement, in violation of the ice rules (Figure 1c). This breaking of the ice rules at low temperature is associated with a metamagnetic phase transition below $T$ ~ 400 mK[23] that closely mimics a liquid-gas transition line, including the existence of a critical point [3, 17, 23, 24].

We measured the real and imaginary parts of the ac magnetic susceptibility ($\chi'$ and $\chi''$) of two different crystals of $Dy_2Ti_2O_7$ grown by the floating zone method and cut to dimensions 0.7 mm x 0.7 mm x 2.2 mm and 0.7 mm x 0.7 mm x 2.1 mm, respectively, with the c axis aligned along the [111] direction. All of the data reported here were taken with the ACMS option of the Quantum Design Physical Property Measurement System [25]. The crystals were aligned with both the ac coil set and the dc field of the cryostat to within ± 2 degrees. The system has a minimum temperature of 1.8 K and a frequency range of $f$ = 10 Hz – 10 kHz, and the data did not depend on the magnitude of the ac field ($H_{ac}$ = 1 Oe for all measurements except where noted otherwise, with linear response). The data from the two crystals, grown by different groups, were consistent and corrected for demagnetization effects [25]. While the demagnetization corrections did shift our determination of the spin relaxation time[26], we note that the correction to the data was otherwise small and did not qualitatively change the results.

In figure 2 we plot the zero field temperature dependence of the ac susceptibility, which shows characteristic behavior seen previously in both powder and single crystal samples [2, 6, 8, 9, 12-16, 20, 25, 27]. The low frequency data show a smooth and monotonic rise in the real part of the susceptibility, $\chi'(T)$ with decreasing temperature. By contrast, at higher frequencies, the data show a spin freezing near 16 K as their dynamics slows down; this has been attributed to the single-ion anisotropy of the Dy moments [2]. There is another maximum observable at lower temperatures for the high frequency data, reflecting the freezing of the moments into the characteristic two-in/two-out local structure of the low temperature spin ice state. Note that at lower frequencies the spin-ice freezing is not apparent, but above approximately $f$ = 1 kHz the freezing starts to become observable around $T$ = 2 K. This higher frequency regime elucidates the development of the spin ice correlations at these relatively high temperatures. The behavior is consistent with the dc magnetization data M(H), at 1.8 K which show the development of spin ice correlations through an incipient [111] magnetization plateau (Figure 2 inset).

In figure 3, we show the ac susceptibility as a function of the applied dc magnetic field. In figure 3a and 3c, we show data taken at 10 kHz at different temperatures. At 7 K, well above the onset of spin ice correlations, the data decrease monotonically with increasing field, reflecting the suppression of susceptibility as the external field clamps the spins along the field direction. As the temperature is lowered, the data show an emergence of distinct regimes (marked by red arrows in the figures) in the field dependence of both $\chi'(H)$ and $\chi''(H)$, the physical origins of which we discuss below. In figure 3b and 3d, we show the same data at $T$ = 1.8 K but at different measurement frequencies. The notable feature in Figure 3a and 3b is the large peak in $\chi'(H)$ near $H$ = 1 T, that corresponds in field to the metamagnetic transition observed below $T$ ~ 0.5 K where the spins are kinetically frozen. While the persistence of the feature to such high temperatures is somewhat surprising, other measurements have indicated considerable spin-ice-like correlations developing in our temperature range[2, 12, 18]. Our observation of the 1 T peak suggests that the understanding of spin ice physics at low temperatures may also be relevant in our temperature range, a theme that we build upon in our analysis below.

The local maximum in the frequency dependent imaginary part of the susceptibility, $\chi''(f)$, at any given temperature and field can be used to define a characteristic spin relaxation time, $\tau$ = $1/f_{max}$[28] (note that this definition is consistent with previous works of our group with regard to the potential inclusion of a factor of $2\pi$ in the definition of frequency). This determination of $\tau$ is based on the Casimir du Pre relation[28], and previous studies have indicated that in this temperature range, the experimental form of $\chi''(f)$ is close to that for an ideal system with a single relaxation time. This is also the case in our system, with relatively little broadening, and we are therefore able to follow the changes in spin relaxation processes as a function of magnetic field based on the data in figure 4a (additional data can be found in the Supplementary Information). The behavior of $\tau(T)$ in zero field is qualitatively consistent with previously published data[16], although the measured values of $\tau$ are notably lower than those measured for polycrystalline samples. The demagnetization correction for our samples is not the cause of this discrepancy, since correcting the powder data would give an even longer relaxation time, and thus we suspect the presence of a slight oxygen stoichiometry difference in powder samples or that the ions near grain surfaces affected those results. As shown in figure 4b, we find a systematic field dependence to $\tau$ with maxima in $\tau(H)$ near 0.25 T and 0.6 T that are suggestive of complex dynamics among the spins.

We now discuss our data and consider how the results can be related to the polarization of the rare earth ions on the kagome and triangular lattice sites. With the application of a magnetic field, the spins on the triangular lattice sites are the first to polarize owing to the larger Zeeman coupling. This presumably accounts for the majority of the susceptibility decrease at low fields, as it is reflected in the sharp rise in magnetization that is seen below H ~ 0.25 T (figure 2 inset). We observe a corresponding sharp decrease in $\chi'(H)$ (figure 3a and 3b) and a sharp increase in $\tau(H)$ (figure 4b) -- as the triangular lattice spins become more strongly enslaved to the applied DC field, they become less responsive to the weak oscillating field. Note that the features in the susceptibility in figures 3a and 3b correspond to the same field values from the separate measurement of the spin relaxation time in figure 4b. While these features are only sharp enough at $T$ = 1.8 K to determine an exact field value, we note that ref.[29] does predict a temperature dependence for the onset of the kagome ice state that is proportional to the external field in the limit of low temperature.

As the field is increased above 0.25 T, $\chi'(H)$ in figure 3a and 3b decreases more slowly, reaches a minimum around 0.6 T, and then rises to a peak near 1.0 T before dropping to zero at the highest fields. In the same regime, $\tau(H)$ in figure 4b decreases from a maximum near 0.25 T down to a first minimum around 0.4 T; it then rises to another maximum near 0.6 T, decreases to a second minimum around 1 T before rising again at the highest fields. The initial decrease in $\tau(H)$ in figure 4b, just above 0.25 T, is consistent with a crossover between regimes where the triangular lattice sites and the kagome lattice sites dominate the susceptibility (making the assumption that the relaxation time for the kagome site spins is increased less by the applied field due to their incomplete alignment with the field). At higher fields, the triangular layer spins are largely saturated as their largest energy scale quickly becomes the Zeeman energy, pinning them along the field. Therefore, it is unlikely that the triangular spins are responsible for the non-monotonic behavior of $\tau(H)$ observed for $H > 0.25$ T. The origin of this behavior presumably arises instead from spin dynamics within the kagome layers.

Let us now consider the dynamics of the spins in the kagome layer in detail for the field regime $H > 0.24$ T. As detailed below, we find that the language of monopole excitations provides a useful framework for considering the data, despite the relatively high temperatures that correspond to a high density of such excitations. The monopole excitation picture has successfully explained much of the low-temperature physics of spin ice, and it allows us to identify the field regimes emerging at high temperature that then persist to become well-developed at low temperatures. However, we note that the parallel presentation of the monopole framework is not required for the interpretation of this physics.

We note that flipping a kagome spin can have two results. Either a monopole is moved between a 'low Zeeman state' tetrahedron (which has all moments aligned with the field) and 'high Zeeman state' one (with two moments pointing against the field); or a monopole-antimonopole pair in adjacent tetrahedra in the same Zeeman state is created. We ignore here any contributions due to double monopoles (4-in or 4-out tetrahedra) as their energy cost is much larger. As the field is increased, the energy difference between monopoles in low vs. high Zeeman states grows, thus reducing the ability of the system to respond to an applied field by moving monopoles between Zeeman states (i.e., $\chi'$ continues to decrease and $\tau(H)$ begins to increase to the second maximum near 0.6 T).

In the low-field region, below $H \sim 0.24$ T where the triangle plane spins dominate the susceptibility, the spin relaxation time $\tau(H)$ increases monotonically as the field slows down the spin dynamics. However, as we consider higher fields (above $H \sim 0.24$ T) where the triangular layer spins are polarized, the field dependence of our measured $\tau(H)$ is dominated by the spins in the kagome planes. Whilst this mechanism is impaired by the interaction energy cost of creating new monopoles, it is in fact facilitated by a competing, and growing, Zeeman energy. Therefore, the ability of the system to respond to fluctuations in the applied [111] field via the creation or annihilation of monopoles increases with the field. This can account for the increase in $\chi'(H)$ in figures 3a and 3b and decrease in $\tau(H)$ in figure 4b -- leading to the minimum in $\tau(H)$ at 1.0 T. Essentially, at this field, the energy cost of flipping a spin is very low since the Zeeman energy and the interaction energies are balanced. At the highest fields, this mechanism is again switched off as the Zeeman energy now dominates and the system is saturated, with all spins pinned by the field, leading to the suppression of $\chi'(H)$ that we observe. We note that $\tau(H = 1$ T$)$ could approximate the spin-flip timescale of the system in a non-interacting setting; however, determining the exact value with our apparatus is hindered by our maximum measurement frequency of $f = 10$ kHz.

We note that the field position of the 1.0 T peak in $\chi'(H)$ in figure 3a has almost no temperature dependence. This is consistent with observations from low temperatures showing that the metamagnetic phase boundary around 0.9 T has little temperature dependence[30]. This can be understood from the Clausius-Clapeyron equation which relates the slope of the phase boundary to the ratio of the entropy difference between the two phases – which despite the extensive degeneracy is still relatively small for kagome ice[21, 22] and the saturated state – and the jump in the magnetization, which is sizeable for the large magnetic moments. As the temperature is increased, the metamagnetic transition terminates in a critical point[30], after which it is replaced by an increasingly broad crossover. This leads to a flattening out of the peak, until thermal fluctuations dominate entirely when the temperature is above the Zeeman energy of a triangular spin at the transition, $T = 10\mu_B \cdot B/k_B \approx 6.72$ K.

Another notable aspect of the 1.0 T peak is the strong frequency dependence that we observe below the peak and the almost complete absence of frequency dependence above the peak (Figure 3b). Frequency dependence in the susceptibility is typically associated with a complex energy landscape[31], and our data thus indicate the energy landscape to be much simpler at fields above the peak than below. This is in keeping with the disordered nature in the highly degenerate kagome ice state, which is in contrast to the ordered saturated state in which each tetrahedron has the same moment arrangement. In the kagome ice state, the small but non-vanishing zero-point entropy[17] (reduced from but analogous to that of the zero-field spin ice state[16]) implies that some local low-energy modes are available, but not uniformly across all the crystallographically equivalent spins of the system. Quite surprising in our data, however, is the persistence of such a physical scenario at such high temperatures where the system is still quite dynamic.

A separate curious feature in figure 3a is the crossing point of all the $\chi'(H)$ data at different temperatures. Such a crossing point indicates that the susceptibility is temperature-independent at that value of magnetic field, and indeed the data in the inset to figure 3c show that both the real and imaginary components of the 10 kHz susceptibility are nearly temperature independent in a field of 0.688 T (variation of less than 10% for $\chi'(T)$ and 7% for $\chi''(T)$ between 12 K and our lowest temperature of 1.8 K). Such temperature independence could simply be associated with a fortuitous cancellation of different effects. The extended temperature range and the coincidence of the onset temperature with that of quantum spin relaxation observed previously, however, both hint at the intriguing possibility that this particular field value is coincident with a purely quantum mechanical process in the higher frequency ranges (see Supplementary Information for data at other frequencies).

In summary, our data allow us to identify the onset of qualitatively different dynamical regimes of spin ice already at high temperatures as the ice rules are only starting to assert themselves. Saliently, the frequency dependence of the kagome ice data evidences its collective nature incorporating the kinematic constraint imposed by the incipient ice rules at these temperatures. Future studies should extend these results down in temperature through the onset of the first-order metamagnetic transition, and extend the frequency range higher to explore the possible quantum relaxation effects in the kagome ice state. We find that the results can be connected to the low-temperature behavior by a qualitative description in terms of emerging monopole excitations in this regime. Somewhat surprisingly, therefore, the insights gained from the monopole picture, appropriate for the low temperature regime of spin ice, appear to also be relevant to understanding central qualitative features of the different field regimes that emerge as the ice rules start asserting themselves at much higher temperatures.

* Note that after submission of the present work, two related studies have been posted by Bovo *et al.* at: http://arxiv.org/abs/1210.0106 and Revell *et al.* at: http://www.nature.com/nphys/journal/vaop/ncurrent/full/nphys2466.html

**Acknowledgements:** M. J. M. and P.S. were supported by NSF grant DMR-1104122. S.A.G. was supported by the Royal Society (UK), ANPCyT (Argentina) and CONICET (Argentina). C.C. was supported in part by the EPSRC Postdoctoral Research Fellowship EP/G049394/1.

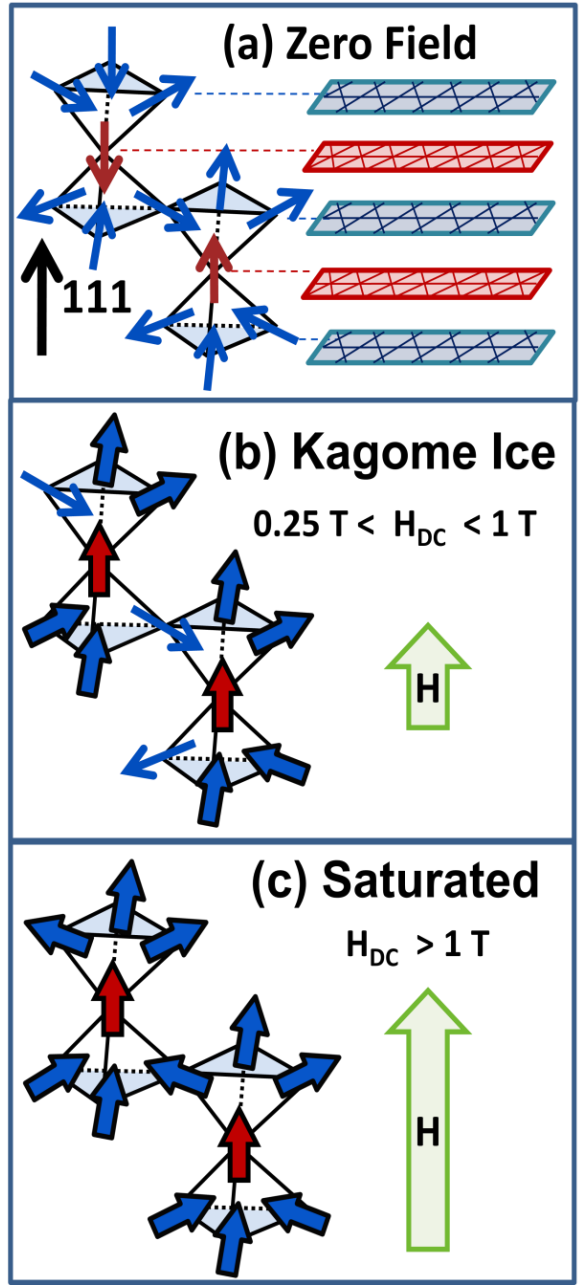

**Figure 1.** Cartoon of the static stages of magnetization under an applied dc field for the [111] direction—alternating layers of kagome and triangular planes. Blue spins lie in the kagome planes and red spins lie in the triangular planes. Bold arrows indicate that the spins are pointing in the direction of the external DC field in the [111] direction. Panel (a) shows the disordered zero-field state with zero net magnetization. Panel (b) shows the partially magnetized ordered (but still 2-in, 2-out) state with one fourth of the spins opposing the external field. Panel (c) shows the fully saturated magnetized state.

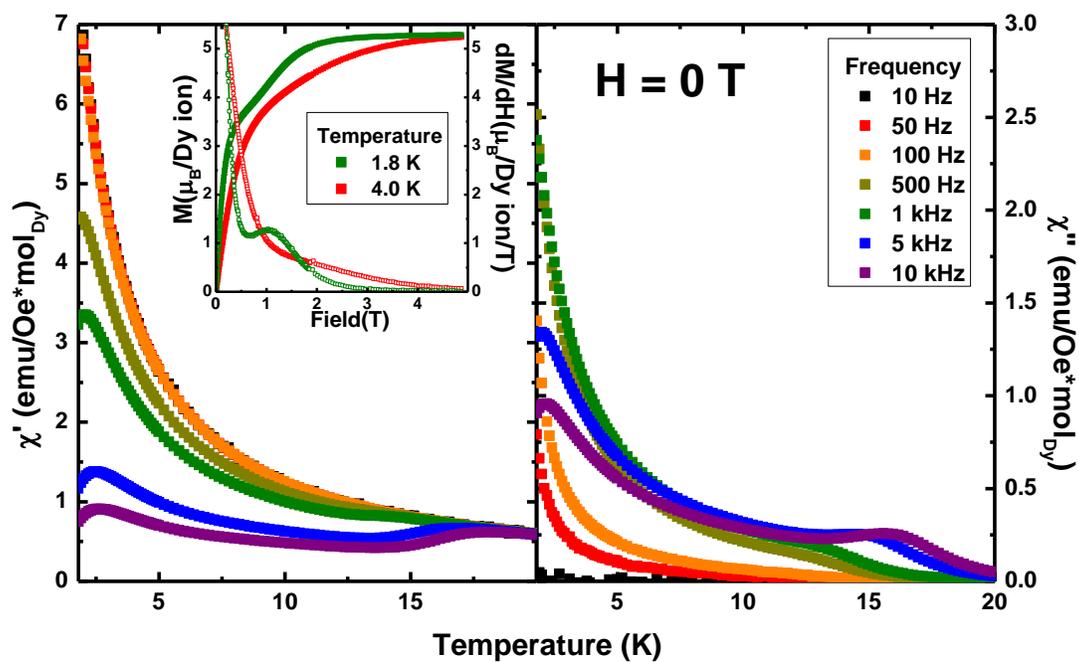

**Figure 2.** Temperature dependent zero field ac susceptibility at different frequencies and (inset) dc magnetization (closed squares) at different temperatures agree well with previously published results[12,16]. The derivative, dM/dH (open squares), is also shown in the inset.

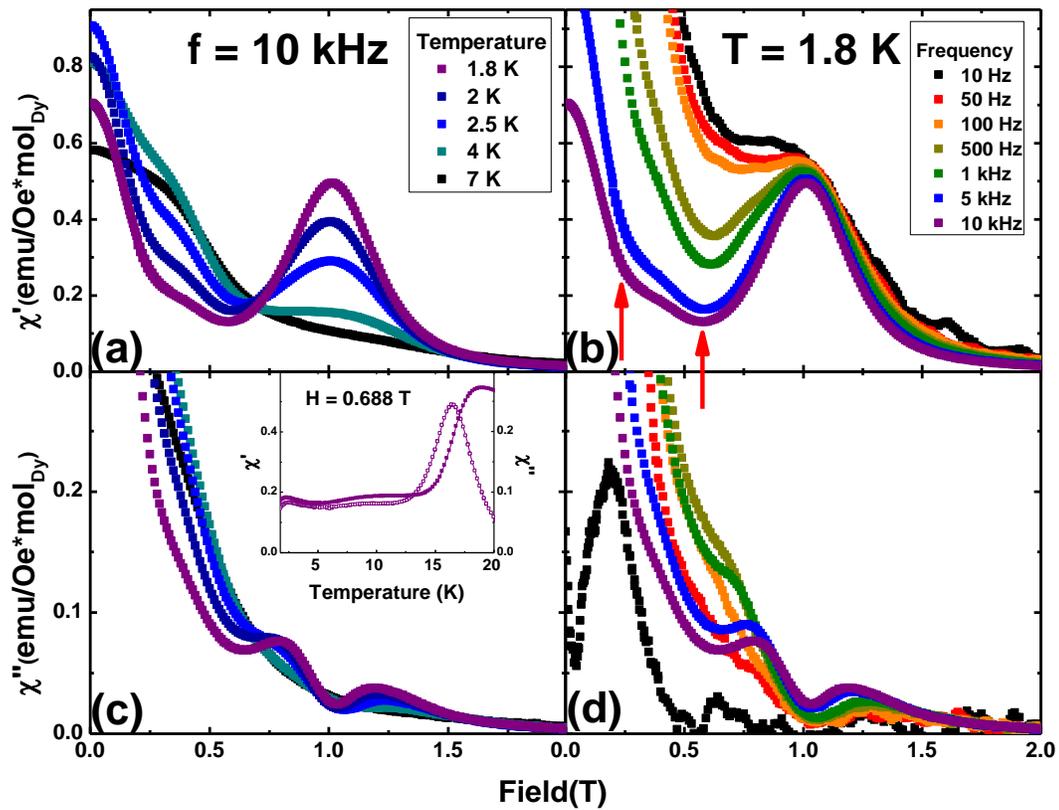

**Figure 3.** Field dependent susceptibilities at (a) 10 kHz, at varying temperatures, and (b) 1.8 K, at varying frequencies show the feature of interest at $H$ = 1 T, along with the corresponding imaginary components (c) and (d), respectively. Inset shows the temperature-independent point at H = 0.688 T in panel (c) for $f$ = 10 kHz.

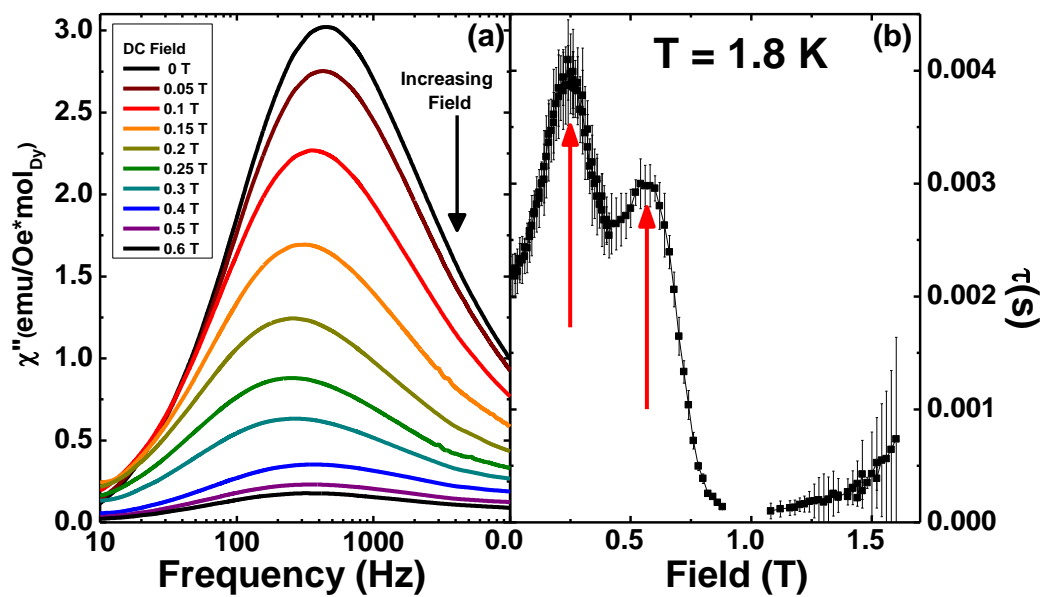

**Figure 4.** (a) Imaginary component of the susceptibility as a function of frequency in different external dc fields (excitation field of 10 Oe). The peak position indicates the characteristic spin relaxation time. (b) A double-peak structure is observed in the characteristic spin relaxation time as an external field is applied, where red arrows correspond to the same field values as in fig. 3(b).

15